\documentstyle[12pt,preprint,aps,floats,epsf,epsfig,amsbsy]{revtex}
\tighten
\newcommand{\bea}{\begin{eqnarray}}
\newcommand{\eea}{\end{eqnarray}}
\newcommand{\beq}{\begin{equation}}
\newcommand{\eeq}{\end{equation}}
\newcommand{\ba}{\begin{array}}
\newcommand{\ea}{\end{array}}
\newcommand{\nn}{\nonumber}
\newcommand{\xc}{X_{c \hspace{-4pt} /}}

\newcommand{\bra}{\langle}
\newcommand{\ket}{\rangle}
\newcommand{\OEightTree}{\bra s g |O_8|b\ket_{\text{tree}}}

\renewcommand{\a}{\alpha}
\renewcommand{\b}{\beta}
\newcommand{\g}{\gamma}

\newcommand{\e}{\epsilon}
\newcommand{\p}{\pi}

\newcommand{\m}{\mu}
\newcommand{\n}{\nu}
\newcommand{\G}{\Gamma}

\renewcommand{\L}{\Lambda}
\renewcommand{\to}{\rightarrow}
\newcommand{\tfrac}[2]{{\textstyle \frac{#1}{#2}}}
\renewcommand{\Re}{\mbox{Re}}
\renewcommand{\Im}{\mbox{Im}}

\begin{document}
\thispagestyle{empty}

\preprint{
\noindent
\hfill
\begin{minipage}[t]{6in}
\begin{flushright}
BUTP--00/21     \\
hep-ph/0008071  \\
\vspace*{1.0cm}
\end{flushright}
\end{minipage}
}

\draft

\title { The rare decay $\boldsymbol{b \to s g}$ beyond leading 
logarithms\footnote{Work partially supported by Schweizerischer
Nationalfonds}}

\author{Christoph Greub and Patrick Liniger} 

\vspace{2.0cm}

\address{
 Institut f\"ur Theoretische Physik, Universit\"at Bern, \\
 CH--3012 Bern, Switzerland} 

\maketitle
\thispagestyle{empty}
\setcounter{page}{0}

\vspace*{1truecm}
\begin{abstract}
We calculate the $O(\alpha_s)$ virtual corrections to the decay width
for $b \to s g$ in the standard model ($g$ denotes a gluon). Also
the corresponding $O(\alpha_s)$ bremsstrahlung corrections to $b \to s g$
are systematically calculated in this paper. The combined result 
is free of infrared and collinear singularities, in accordance with
the KLN theorem. Taking into account the existing
next-to-leading logarithmic (NLL) result for the Wilson coefficient
$C_8^{\rm{eff}}$, a complete NLL result for the branching ratio
${\cal B}^{\rm{NLL}}(b \to s g)$ is derived. Numerically, we obtain
${\cal B}^{\rm{NLL}}=(5.0 \pm 1.0) \times 10^{-3}$, which is more than a factor
of two larger than the leading logarithmic result 
${\cal B}^{\rm{LL}}=(2.2 \pm 0.8) \times 10^{-3}$.
\end{abstract}

\vfill
%July 2000
%\today

%\pacs{Preliminary Version}

\setlength{\parskip}{1.2ex}

\section{Introduction}
\label{intro}
The theoretical predictions for inclusive decay rates of $B$-mesons
rest on solid gounds due to the fact that these rates
can be sytematically expanded in powers of $(\L_{\rm QCD}/m_b)$
\cite{Bigi1,Bigi2}, where the leading term corresponds to the decay width
of the underlying $b$-quark decay. 
As the power corrections  start at $O(\L_{\rm{QCD}}/m_b)^2$ only,
they affect these rates by at most 5\%. Thus 
the accuracy of the theoretical predictions is mainly controlled 
by our knowledge
of the perturbative corrections to the free quark decay.

The charmless inclusive decays, $B \to \xc$, where $\xc$ denotes any hadronic 
charmless final state, are an interesting subclass of the decays mentioned 
above. At the quark level, there are decay modes with three-body final states,
viz. $b \to q' \overline{q}' q$, ($q'=u,d,s$; $q=d,s$) and the modes
$b \to q g$, with two-body final state topology, which contribute 
to the charmless decay width at leading logarithmic (LL) accuracy.
Next-to-leading logarithmic (NLL) corrections to the three-body decay modes
were started already some time ago in ref. \cite{Altarelli}, where
radiative corrections to the current-current diagrams of the operators
$O_1$ and $O_2$ were calculated, together with NLL corrections to the
Wilson coefficients. Later, Lenz et al. \cite{Lenz1,Lenz2} 
included the contributions of the penguin diagrams associated with
the four-Fermi operators $O_1,...,O_6$; 
the effects of the chromomagnetic operator $O_8$ were taken
into account to the relevant precision needed for a NLL calculation. 
Up to contributions from current-current type corrections to the penguin
operators, the NLL calculation for the three quark final states is 
complete. 

In the numerical evaluations of the charmless decay rate, 
the two body decay modes $b \to q g$
were added in refs. \cite{Lenz1,Lenz2} at the LL precision, 
as the full NLL predictions were missing.
It is exactly this gap which we try to fill in the present letter.
We will present the results of a calculation for the 
branching ratio ${\cal B}(b \to s g)$ where NLL corrections are 
systematically included. This implies that 
besides virtual corrections to $b \to s g$
also the process $b \to s g g$
has to be taken into account, as it gives contributions at the same order
in perturbation theory. 
The LL prediction for the branching for $b \to s g$
is known to be ${\cal B}(b \to s g) \approx 0.2\%$ \cite{Ciuchini}; also
the process $b \to s g g$ has been studied in the literature
\cite{Hou,Simma}. In \cite{Simma} a complete calculation was performed
in regions of the phase space which are free of collinear an infrared 
singularities, leading to a branching ratio for $b \to s g g$
of the order of $10^{-3}$.
A complete NLL calculation requires, however, a regularized version
for the decay width $\G(b \to s g g)$ in which infrared and collinear
singularities become manifest. Only after adding the virtually corrected
decay width $\G(b \to s g)$ a meaningful physical result is obtained.
In addition, as we will see later, also the 
tree-level contribution of the operator $O_8$ to the decays 
$b \to s f \overline{f}$, with $f=u,d,s$, has to be included.

The decay $b \to s g$ gained a lot of attention in the last years. 
For a long
time the theoretical predictions for both, the inclusive
semileptonic branching ratio and the charm multiplicity $n_c$ in $B$-meson
decays were considerably higher than the experimental values
\cite{Bigi_Falk}.
An attractive
hypothesis, which would move the theoretical predicitions
for both observables into the direction favored by the experiments,
assumed the rare decay mode $b \to s g$ to be enhanced by
new physics.

After the inclusion of the complete NLL corrections to the decay modes
$b \to c \overline{u} q$ and $b \to c \overline{c} q$ ($q=d,s$) \cite{Bagan},
both the CLEO- and the LEP-data \cite{Japantalk}
are now in agreement with theory \cite{Bagan,Sachrajda},
if one allows the renormalization scale $\mu$ to become as low as $m_b/4$.
 We would like to stress that there is still some room for 
enhanced $b \to s g$,
in particular when using higher values for the renormalization scale.
For theoretical motivations of enhanced $b \to s g$, see ref. \cite{Kaganth}
and references therein. 

We also would like to mention that the component $b \to s g$
of the charmless hadronic decays is expected to manifest itself in 
kaons with high momenta (of order $m_b/2$), due to its two body nature
\cite{Rathsman}. Some indications into this direction were reported by
the SLD collaboration \cite{SLD}. For overviews on
enhanced $b \to s g$, see e.g. refs. \cite{Kagan_hawaii,Neubert98}.   

The remainder of this letter is organized as follows: In section 
\ref{effective_Ham}, we briefly review the theoretical framework.
Section \ref{virtcorr} is devoted 
to the calculation of the virtual corrections
to the decay width for $b \to s g$, while section \ref{bremsmat} deals
with the calculation of the bremsstrahlung corrections to $b \to s g$.
In the short section \ref{Quarkstrahlung} in between, the decay width
for the tree level processes $b \to s f \overline{f}$ mentioned above, 
is given.
In section \ref{brnll} we give the expressions for the NLL branching 
ratio ${\cal B}^{\rm{NLL}}(b \to s g)$, which combines the processes
$b \to s g$, $b \to s g g$ and $b \to s f \overline{f}$. Finally, in section
\ref{numres} the numerical results for ${\cal B}^{\rm{NLL}}(b \to s g)$
are presented. 

\section{Theoretical framework}
\label{effective_Ham}
\noindent 
The analysis of the decays $b \to s g$ and $b \to s g g$ starts
with introducing the effective Hamiltonian
\beq
\label{heff}
{\cal H}_{\rm eff} = - \frac{4 G_F}{\sqrt{2}} \,V_{ts}^* V_{tb} 
   \sum_{i=1}^8 C_i(\mu) O_i(\mu).  
\eeq
where $V_{ij}$ are elements of the CKM matrix, $O_i(\mu)$ are
the relevent operators and $C_i(\mu)$ are the corresponding
Wilson coefficients. 
The full set of operators needed for our application, can be seen
in ref. \cite{Misiak97}. As the Wilson coefficients of the gluonic
penguin operators are small (see eq. (30) in ref. \cite{Misiak97}), 
we neglect them when calculating radiative corrections; we therefore
only list the explicit form of the operators $O_1$,
$O_2$ and $O_8$:
\beq
\begin{array}{llll}
O_1 \,= &\!
 (\overline{s}_L \gamma_\mu T^A c_L)\, 
 (\overline{c}_L \gamma^\mu T^A b_L)\,, 
               &  \quad 
O_2 \,= &\!
 (\overline{s}_L \gamma_\mu c_L)\, 
 (\overline{c}_L \gamma^\mu b_L)\,,   \\[1.002ex]
O_8 \,= &\!
  \frac{g_s}{16\pi^2} \,{\overline m}_b(\mu) \,
 (\overline{s}_L \sigma^{\mu\nu} T^A b_R)
     \, G^A_{\mu\nu} \ .
\end{array} 
\label{opbasis}
\eeq
Here $T^A$ stand  for the $SU(3)_{\rm{color}}$ generators.
The small CKM matrix element $V_{ub}$ as well as the $s$-quark mass
are also neglected. 

It is well-known that in this formalism the large QCD logarithms,
present in the decay amplitude for $b \to s g$, are contained in 
resummed form in the Wilson coefficients when choosing the renormalization
scale $\mu$ at the order of $m_b$. The LL (NLL) Wilson coefficients
contain all terms of the form $\a_s^n \ln^n(m_b/M)$
 $(\a_s \, \a_s^n \ln^n(m_b/M))$, where $M=m_t$ or $m_W$
and $n=0,1,2,\ldots$.

The LL prediction for the decay amplitude for $b \to s g$ is then obtained
by calculating the matrix elements $\bra s g |O_i| b \ket$ at order
$g_s$ and weighting them with the leading logarithmic Wilson coefficients.
In the naive dimensional regularization (NDR) scheme which we use in this
paper, there are one-loop contributions of order $g_s$ for $i=3,4,5,6$
and the tree-level contribution of $O_8$.
The effect of the matrix elements for $i=3,4,5,6$ can be absorbed into
the effective Wilson coefficient (see \cite{Misiak97}) 
$C_8^{\rm{eff}} =  C_8 + C_3 - \tfrac{1}{6} \, C_4 + 20 \, C_5 - 
\tfrac{10}{3} \, C_6$. 

The NLL corrections for the decay amplitude for $b \to s g$ receives
two contributions: The first one arises when combining the lowest
order matrix elements (of order $g_s$) with the NLL Wilson coefficient,
while the second one arises when calculating explicit order $\a_s$ corrections
to the matrix elements of the operators which are then weighted 
with LL Wilson
coefficients. As the operators $O_1$ and $O_2$ have vanishing matrix elements
for $b \to s g$ at order $g_s$ and the NLL corrections 
connected with the operators
$O_3,...,O_6$ are neglected, the only Wilson coefficient needed to NLL
precision is that of the operator $O_8$. The necessary ingredients,
the anomalous dimension matrix to $O(\a_s^2)$ and 
%have been worked out
%in ref. \cite{Misiak97}, while 
the $O(\a_s)$ matching condition
for the operator $O_8$ were given in refs. \cite{Misiak97} and 
\cite{Adel}, respectively. 
A practical parametrization 
for $C_8^{\rm{eff}}(\mu)$  will be given in ref. \cite{GLprep}.
A rather complete list of references on 
matching conditions, anomalous dimension matrices, and 
on the process $b \to s \gamma$, which is similar in many respects
to $b \to s g$, is given e.g. in ref. \cite{BG98}. 
%%%%%%%%%%%%%%%%%%%%%% SECTION 3 %%%%%%%%%%%%%%%%%%%%%%%%%%%%%%%%
\section{Virtual corrections to $\boldsymbol{O_1}$, $\boldsymbol{O_2}$
 and $\boldsymbol{O_8}$}
\label{virtcorr}
\subsection{Virtual corrections to  $\boldsymbol{O_1}$ and $\boldsymbol{O_2}$}
\begin{figure}[t]
\begin{center}
\leavevmode
\includegraphics[width=\textwidth]{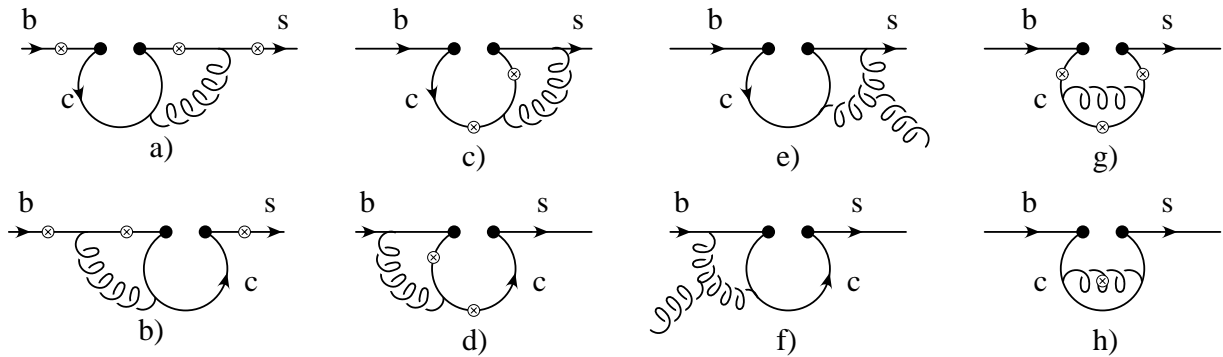}
\vspace{3ex}
\caption[f1]{Complete List of two-loop Feynman diagrams for $b \to s g$
associated with the operators $O_1$ and $O_2$.
The fermions ($b$, $s$ and $c$ quarks) are represented by solid lines;
the wavy  lines represent gluons.
The crosses denote the possible locations where the gluon is emitted.}
\label{fig:1}
\end{center}
\end{figure}

As the one-loop matrix elements of the operators $O_1$ and $O_2$ 
vanish, we immediately turn to the
two-loop contributions. A complete list of Feynman diagrams
for the matrix elements $\bra s g|O_i| b\ket$ ($i=1,2$) is shown
in fig. \ref{fig:1}.
The diagrams in fig. \ref{fig:1} a), b), c) and d),
in which the emitted gluon is replaced by a photon are the relevant
diagrams
for $b \to s \g$; these were calculated in ref. \cite{GHW}. 
The three possible
diagrams in fig. \ref{fig:1} g) cancel each other when considering the
process $b \to s \gamma$; however, they give a non-vanishing 
contribution\footnote{We thank M. 
Neubert for making us aware of these diagrams.}
to $b \to s g$. There is also another difference: 
while each of the figures a), b), c) and d)
forms a gauge invariant subset in $b \to s \g$, this is no 
longer true for $b \to s g$; a gauge invariant result is only obtained
when all the diagrams in fig. \ref{fig:1} are summed.

The various two-loop integrals are calculated by the  standard 
Feynman parameter technique. 
The heart of our procedure,
which is explained in detail in ref. \cite{GHW} for one of the diagrams
in fig. \ref{fig:1} a),
is to
represent the rather complicated denominators in the  Feynman 
parameter integrals as complex Mellin-Barnes integrals
\cite{Mellin}.
After inserting this
representation and interchanging the order of integration, the
Feynman parameter integrals are reduced to well-known Euler
Beta-functions. Finally, the residue theorem
allows  to represent the remaining
complex integal as the sum over the residues taken at
the pole positions of
Beta- and
Gamma-functions; for a generic diagram $G$, these steps naturally
lead to an expansion in the ratio $z=(m_c/m_b)^2$ of the form
\beq
\label{generic}
G = c_0 + \sum_{n,m} c_{nm} z^n
L^m  \quad ;  \quad z = \frac{m_c^2}{m_b^2} \quad ; \quad L=\ln z \ ,
\eeq
where the coefficients $c_0$ and $c_{n m}$ are independent
of $z$.
The power $n$ in eq. (\ref{generic})
is in general a natural multiple of $1/2$
and $m$ is a natural number including 0. In the explicit
calculation, the lowest $n$ turns out to be $n=1$.
This implies the important fact
 that the limit $m_c \to 0$ exists.
As was shown in \cite{GHW}, the power
$m$ of the logarithm is bounded by  $4$,
independently of the value of $n$.
In our results, which we will present below, all terms up to $n=3$
are retained.

We first present the final result for the 
dimensionally regularized matrix element
$M_2=\bra s g|O_2|b\ket$
which represents the sum of all two-loop diagrams in 
fig. \ref{fig:1}:
\beq
\label{eq:virtualO2}
  M_2     =  \frac{\alpha_s}{4\pi} \OEightTree  \left[
           - \frac{16}{27\,\e} \left( \frac{m_b}{\mu} \right)^{-4 \epsilon}
    +r_2 \, \right] \ ,
\eeq
where the real- and imaginary part of $r_2$ read
\bea
\label{rer2ndr}
\Re( r_2 ) & = &  \frac{1}{648}
                          \big\{ -2170 - 54 \pi^2 + z [48816 - 252 \pi^2 + 
                          (22680
                         - 1620 \pi^2) L \nonumber \\
                   &   & \quad + 2808 L^2 + 612 L^3 - 6480 \zeta(3)] \nonumber
                          \\
                   &   & \quad -12672 \pi^2 z^{3/2}+z^2 [66339 + 1872 \pi^2 +
                          (-40446 + 1512 \pi^2) L \nonumber \\ 
                   &   & \quad + 6642 L^2 - 1008 L^3 + 7776
                          \zeta(3)] \nonumber \\
                   &   & \quad +z^3 [-3420 - 60 \pi^2 - 6456 L + 7884
                          L^2] \big\} \nn \\
\Im( r_2 ) & = &  \frac{\p}{27}
                       \big\{
                  -28+z [549 - 24 \pi^2 + 153 L + 72
                         L^2] \nonumber \\ 
                   &   & \quad +z^2 [-432 + 30 \pi^2 + 54 L - 90 L^2]+z^3 [-259 +
                         192 L]  \big\} \ .
\eea
The symbol $\zeta$ stands for the Riemann
Zeta function, with $\zeta(3) \approx 1.2021$. Finally,
the quantity $\OEightTree$ denotes the tree level
matrix element of the operator $O_8$.
%As such, it contains the 
% running $b$-quark mass and the running strong coupling constant,
%both evaluated at the scale $\mu$ (see eq. (\ref{operators})).
%However, as the corrections to $O_2$ are explicitly proportional
%to $\a_s$, we are allowed (modulo higher order terms) to identify
%the running $b$-quark mass with the pole mass $m_b$; in the same spirit
%we can identify the strong coupling constant with $g_s(m_b)$. 
%With this interpretation, which we will use in the
%following, $\OEightTree$ is a scale independent quantity.
%, reading
%\beq
%\label{o8tree}
%\OEightTree = m_b \, \frac{g_s(m_b)}{8\p^2} \,
%\bar{u}(p') \, \epsl \qsl R \,T^A \, u(p) \quad .
%\eeq

To obtain the renormalized matrix element $M_2^{\rm{ren}}$ 
associated with the operator
$O_2$, the corresponding counterterms have to be included.
This means that we have to take into account
the one-loop matrix elements of the four Fermi
operators
$\delta Z_{2j} O_j$ ($j=1,...,6$) and the tree level contribution 
of the magnetic operator $\delta Z_{28} O_8$.
In the NDR scheme  the
only non-vanishing contributions
to $b \to s g$ come from  $j=4,8$.
The operator renormalization
constants $Z_{ij}$
can be extracted from the literature \cite{Misiak97}
in the context of the leading
order anomalous dimension matrix. 
%\footnote{Note that the
%effective anomalous dimension matrix $\g^{0,\rm{eff}}$ given
%in \ref{Misiak97} has to be converted into $\g^{0}$, before the relevant
%$\d Z$-factors can be read off.}. 
One obtains the counterterm contribution
%The entries needed for the renormalization of the $O_2$
%contribution to the amplitude for $b \to s g$ read:
%\beq
%\label{zfactors2}
%\delta Z_{24} =  \frac{\a_s}{6 \pi \e}  \quad , \quad
%\delta Z_{28} =   \frac{19 \, \a_s}{108 \pi \e}  \quad .
%\eeq
%The counterterm contribution $M_2^{\rm{ct}}$ is then given by
%
\beq
\label{ct2}
M_2^{\rm{ct}} = \bra s g |\delta Z_{24} O_4+\delta Z_{28} O_8|b \ket
= \left( -\frac{\a_s}{36 \pi} \,
  \frac{1}{\e} \left(
\frac{m_b}{\mu} \right)^{-2\e} +
\frac{\a_s}{ \pi} \, \frac{19}{108}
\, \frac{1}{\e} \right) \OEightTree \quad .
\eeq
We note that there are no one-loop contributions to 
the matrix element for $b \to s g$ from counterterms
proportional to the evancesent operator 
$P_{12}$ given in
appendix A of ref. \cite{Misiak97}.
Adding the regularized two-loop result in eq. (\ref{eq:virtualO2})
and the counterterm in
eq. (\ref{ct2}), we find the renormalized result
for $M_2$ in the NDR scheme:
\beq
\label{m2lr}
M_2^{\rm{ren}} = \OEightTree \, \frac{\a_s}{4 \p} \,
\left( \ell_2 \ln \frac{m_b}{\mu}  + r_2 \right) \quad ,
\eeq
where
$r_2$ is given in eq. (\ref{rer2ndr}) and $\ell_2 = 70/27$.

By doing analogous steps, we obtain the renormalized version of 
$M_1=\bra sg|O_1|b\ket$: 
\beq
\label{m1lr}
M_1^{\rm{ren}} = \OEightTree \, \frac{\a_s}{4 \p} \,
\left( \ell_1 \ln \frac{m_b}{\mu}  + r_1 \right) \quad ,
\eeq
with $\ell_1=173/162$ and
\bea
\label{rer1ndr}
\Re( r_1 )  & = &  -\frac{1}{3888}
                          \big\{ 4877 - 54 \pi^2 + 36 z [1086 + 29 \pi^2 + (360
                         + 36 \pi^2) L \nonumber \\
                   &   & \quad + 51 L^2 + 8 L^3 + 144 \zeta(3)] \nonumber
                          \\
                   &   & \quad -12672 \pi^2 z^{3/2}+ 9 z^2 [6615 - 80 \pi^2 +
                          (-4494 + 384 \pi^2) L \nonumber \\ 
                   &   & \quad + 864 L^2 - 148 L^3 + 864
                          \zeta(3)] \nonumber \\
                   &   & \quad +12 z^3 [93 + 76 \pi^2 - 1186 L + 900
                          L^2] \big\} \nn \\
\Im( r_1 ) & = &  -\frac{\pi}{324}
                       \big\{
                       25 +6 z [75 + \pi^2 + 24 L -3  L^2] \nonumber \\ 
         &   & \quad +6 z^2 [-171 + 19 \pi^2 + 72 L - 57 L^2]+ 2 z^3 [-421 +
                         192 L]  \big\}
\eea
For  $z \ge 1/4$ the imaginary parts of $r_1$ and $r_2$ must
vanish exactly; 
our results fulfill this property to high accuracy when
retaining terms up to $z^3$ in the expansion.
\subsection{Virtual corrections to $\boldsymbol{O_8}$}
\label{virto8}
We present now the results of  the virtual 
corrections to the matrix element
$M_8 = \bra s g |O_8|b\ket$.
\begin{figure}[t]
\begin{center}
\leavevmode
\includegraphics[width=\textwidth]{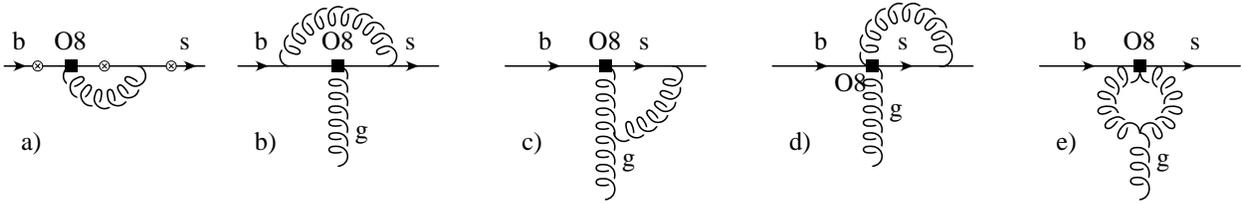}
\vspace{3ex}
\caption[f1]{Complete list of Feynman diagrams associated with the 
operator $O_8$. The analogues of a), c) and d), where the
virtual gluon hits the $b$-quark instead of the $s$-quark,
are not shown explicitly.
The real gluon can
be attached to any of the crosses shown in a).}
\label{fig:2}
\end{center}
\end{figure}
As the contributing Feynman graphs in fig. 
\ref{fig:2} are all one loop diagrams,
the computation of $M_8$ is  straightforward.
We use dimensional regularization for both, the ultraviolet and 
the infrared singularities. Singularities which appear in the situation
where the virtual gluon becomes almost real and collinear with the
emitted gluon are also regulated dimensionally; on the other hand,
those singularities where the almost real internal gluon
is collinear with the $s$-quark, are
regulated with a small strange quark mass $m_s$;
the latter manifest themselves in logarithmic terms
of the form $\ln(\rho)$, where $\rho=(m_s/m_b)^2$.
 
We were able to separate the ultraviolet $1/\e$ poles from those
which are of infrared (and/or collinear) origin. For ultraviolet poles
we use the symbol $1/\e$ in the following, while collinear and infrared
poles are denoted by $1/\e_{\rm IR}$.

%When working in Feynman gauge for the gluon propagator, the individual
%diagrams contributing to $M_8$ have the following infrared/collinear
%properties: 
%a) the diagrams in fig.~\ref{figo8:1new}  are free 
%of infrared and collinear singularities; 
%b) the diagram in fig.~\ref{figo8:2new}
%has a combined collinear and infrared singularity of the form
%$\ln(\rho)/\e_{\rm IR}$ as well as collinear singularities of the form 
%$\ln^2(\rho)$ and $\ln(\rho)$; 
%c) the diagram in fig.~\ref{figo8:3new} (on the left) has
%combined infrared/collinear singularities of the form $1/\e^2_{\rm IR}$
%or  $\ln(\rho)/\e_{\rm IR}$ as well as $1/\e_{\rm IR}$ poles. 
%d) the diagram in fig.~\ref{figo8:3new} (right) has
%combined infrared/collinear singularities of the form $1/\e^2_{\rm IR}$
%as well as $1/\e_{\rm IR}$ poles.
%e) the diagram
% in fig.~\ref{figo8:4new} (left) has a collinear singularity of the
%form $\ln(\rho)$, while the diagram on the right is free of
%infrared and collinear singularities; f) finally, the diagram
%in fig. \ref{figo8:5new} has an infrared singularity of the form
%$1/\e_{\rm IR}$; more precisely, this diagram is proportional to the
%combination $(1/\e-1/\e_{\rm IR})$.
 
As the results of the individual diagrams are not very instructive,
we only give their sum:
\beq
\label{o8result}
M_8 = \frac{\a_s}{4\p} \, f_8 \,  \OEightTree \ ,
\eeq
with
\bea
\label{f8}
f_8 &=& \left[ \,- \frac{3}{\e_{\rm IR}^2} - \frac{(4 \ln(\rho)+9+9i\p)}{3\e_{\rm IR}} +
 \frac{11}{3\e} \right] \, \left( \frac{m_b}{\mu} \right)^{-2\e} + 
 \nn \\
& & \quad \frac{1}{3} \, \left[ \frac{59\p^2}{12} + 1
     -8 \ln(\rho) + 2 \ln^2(\rho)-8i\p \right] \ .
\eea
%We would like to mention that we did not include diagrams with
%self energy insertions in the external legs. As we work in an on-shell
%renormalization scheme with respect to quark and gluon fields, such diagrams
%are cancelled against counterterm contributions.
%
An ultraviolet finite result is obtained by adding the contribution
from the counterterm which is 
generated by expressing the bare
quantities in the tree-level matrix element of $O_8$
by their renormalized counterparts. It has the structure
\beq
\label{m8ct}
M_8^{\rm{ct}} = \delta R \, \OEightTree \quad ,
\eeq
where the factor $\delta R$ is given by
$\delta R = \sqrt{Z_2(m_b)} \, \sqrt{Z_2(m_s)} \, \sqrt{Z_3} \, Z_{g_s} \,
Z_{m_b} \, Z_{88} \, - \,  1 \ $.

$Z_2(m_b)$, $Z_2(m_s)$ and $Z_3$ denote the on-shell 
wave function renormalization factors
of the $b$-quark, the $s$-quark and the gluon, respectively.
$Z_{g_s}$ and $Z_{m_b}$ denote the $\overline{\mbox{MS}}$ 
renormalization constants for
the strong coupling constant $g_s$ and the $b$-quark mass, which
appear explicitly in the definition of the operators (see eq.
(\ref{opbasis})).  
Finally, $Z_{88}$ is the renormalization factor of the operator $O_8$.
The $Z$-factors of the fields, of the masses and
of the strong coupling constant are given in text books, while
$Z_{88}$ can be extracted from the anomalous dimension matrix
in ref. \cite{Misiak97}; we therefore 
immediately give the expression for  $\delta R$: 
\beq
\label{Rexplizit}\
\delta R = -\frac{\a_s}{4\p} \, \left[
\frac{11}{3\e} + \frac{31}{6\e_{\rm IR}} - 8 \ln \frac{m_b}{\m}
-\frac{2}{3} \sum_f \ln \frac{m_f}{\m} + \frac{16}{3} - 2 \ln\rho
\right]\ .
\eeq
The term $\sum_f$ originates from fermion self-energy diagrams
contributing to the on-shell renormalization constant $Z_3$
of the gluon field; $f$ runs over the five flavors 
$u$, $d$, $s$, $c$ and $b$. 

When adding the regularized matrix element of $O_8$ in eq. 
(\ref{o8result}) and
the counterterm contribution $M_8^{\rm{ct}}$ in eq. (\ref{m8ct}),
we obtain the renormalized result
\beq
\label{m8ren}
M_8^{\rm{ren}} = \frac{\a_s}{4\p} \, f_8^{\rm{ren}} \, \OEightTree \ ,
\eeq
with
\bea
\label{f8ren}
f_8^{\rm{ren}} &=& \left[ \, - \frac{3}{\e_{\rm IR}^2} - 
     \frac{(8 \ln(\rho)+49+18i\p)}{6\e_{\rm IR}} \right] \,
  \left( \frac{m_b}{\mu} \right)^{-2\e} -
\frac{29}{3} \ln \frac{m_b}{\m} \nn \\
& & \quad +\frac{2}{3} \sum_f \ln \frac{m_f}{\m} \,
     -5 + \frac{59\p^2}{36} -\frac{2}{3} \ln\rho+ \frac{2}{3} \ln^2 \rho
     - \frac{8}{3} i \p \ .
\eea
We anticipate that the singular terms of the form $1/\e^2_{\rm IR}$, 
$1/\e_{\rm IR}$ 
and $\ln \rho$ in eq. (\ref{f8ren}) will cancel 
%(at the level of the decay widths) 
against the corresponding singularities 
in the result for the gluon bremsstrahlung corrections to $b \to s g$. On the
other hand, the terms
$\ln(m_f/\m)$, which also represent some kind of singularities
for the light flavor $f=u,d,s$ are not cancelled in this way.
Keeping in mind that they originate from the 
fermionic contribution to the 
renormalization 
factor $Z_3$,
it is expected that they will cancel against
the logarithms present in the decay rate $\G(b \to s f \overline{f})$ 
($f=u,d,s$). 
%To cancel these unphysical terms, we will  include 
%the $O_8$ contribution 
%to this process in section \ref{Quarkstrahlung}.
%   
\subsection{Virtual corrections to the decay width
 $\boldsymbol{\G(\symbol{98} \to \symbol{115} \symbol{103} })$}
\label{Gammavirt}
We are now in the position to write down the renormalized version of 
the matrix $M^{\rm{ren}}(b \to s g)$ 
element for $b \to s g$, where the virtual order
$\a_s$ corrections are included. We obtain:
\bea
\label{matelvirt}
M^{\rm{ren}}(b \to s g) &=& \frac{4G_F i }{\sqrt{2}} \, V_{ts}^* V_{tb} \,
\bigg\{ C_8^{\rm{eff}}  + 
\frac{\a_s}{4\p} \bigg[ C_1^0 (\ell_1 \ln \frac{m_b}{\mu}+r_1) +
                        C_2^0 (\ell_2 \ln \frac{m_b}{\mu}+r_2) +
 \nn \\
& &  \quad
       C_8^{0,\rm{eff}} \, f_8^{\rm{ren}} \, \bigg]
\bigg\} \, \bra sg|O_8(\mu)|b \ket_{\rm{tree}}  .
\eea
The quantities $r_1$, $r_2$ and $f_8^{\rm{ren}}$ 
are given in eqs. (\ref{rer1ndr}),  
(\ref{rer2ndr}) and (\ref{f8ren}), respectively. 
As eq. 
(\ref{matelvirt}) shows, 
$C_8^{\rm eff}$ is the only Wilson coefficient needed to NLL precision.
For the following it is useful to decompose it as
\beq
\label{c8decomp}
C_8^{\rm{eff}} = C_8^{0,\rm{eff}} + \frac{\a_s}{4\p} \, C_8^{1,\rm{eff}} \ . 
\eeq
The symbol
$\bra s g|O_8(\mu)|b \ket_{\rm{tree}}$ in eq. (\ref{matelvirt})
denotes the tree level matrix
element of $O_8(\mu)$, which contains the running $b$-quark mass
and the strong running coupling constant at the scale $\mu$.  
In order to get expressions where the $b$-quark
mass enters as the pole mass, and the strong coupling constant enters
as $g_s(m_b)$, we rewrite       
$\bra s g|O_8(\mu)|b \ket_{\rm{tree}}$ as
\beq
\label{treerel}
\bra s g|O_8(\mu)|b \ket_{\rm{tree}} =
\bra s g|O_8|b \ket_{\rm{tree}} \, \left[
1 + \frac{2\a_s}{\p} \ln \frac{m_b}{\m} - \frac{4}{3} \frac{\a_s}{\p}
+\frac{\a_s}{4\p} \, \b_0 \, \ln \frac{m_b}{\m} \right] \ ; \quad
\b_0=\frac{23}{3}  \ .
\eeq
The symbol $\bra s g|O_8|b \ket_{\rm{tree}}$ then stands for
the tree level matrix element of $O_8$ in which $\overline{m}_b(\m)$
and $g_s$ have to
to be indentified with the pole mass $m_b$ and $g_s(m_b)$, respectively.
After inserting 
eqs. (\ref{c8decomp}) and (\ref{treerel}) into eq. 
(\ref{matelvirt}), the corresponding decay width $\G^{\rm{virt}}$ 
is obtained in the standard way. One gets:
\bea
\label{gammavirt}
\G^{\rm{virt}} &=&  
 \frac{\a_s(m_b) \, m_b^5 }{24\p^4} |G_F V_{ts}^* V_{tb}|^2 \,
\left\{ \left( C_8^{0,\rm{eff}} \right)^2 + 
\frac{\a_s}{4\p} \, C_8^{0,\rm{eff}} \,\left[
2 \, C_8^{1,\rm{eff}} 
+2(8+\b_0) \ln \frac{m_b}{\m} \, C_8^{0,\rm{eff}} \right . \right. \nn \\
& &
-\frac{32}{3} C_8^{0,\rm{eff}} 
+ 2 \, C_1^0 
(\ell_1 \, \ln \frac{m_b}{\m} + \Re(r_1)) +
2 \, C_2^0 (\ell_2 \, \ln \frac{m_b}{\m} +\Re(r_2))  \nn \\ 
& & \left. \left. \quad + 2 \, C_8^{0,\rm{eff}} \Re(f_8^{\rm{ren}}) \, (1-\e)
\, \left( \frac{m_b}{\mu} \right)^{-2\e} \,
\left( 1+2\e-\frac{1}{4}(\pi^2-16)\e^2 \right) \right] \right\}  \ .
\eea
We note that due to the infrared poles present in $f_8^{\rm{ren}}$ 
the phase space integrations were done consistently in $d=4-2\e$
dimensions.
% which leads to the last two extra factors in the last term 
%in eq. (\ref{gammavirt}). The other factor, $(1-\e)$, in the
%last term in eq. (\ref{gammavirt}), is due to the fact that
%all the $(d-2)$ possible transverse polarizations of the emitted gluon
%were taken into account.

\section{$\boldsymbol{O_8}$ contribution to the decay width 
 $\boldsymbol{\G(\symbol{98} \to \symbol{115} \symbol{102} \bar{\symbol{102}}})$}
\label{Quarkstrahlung}
As discussed at the end of section \ref{virto8}, we should take into 
account the contribution of the operator $O_8$ to the process
$b \to s f \overline{f}$ $(f=u,d,s)$, in order to cancel the unphysical logarithms
of the form $\ln(m_f/\m)$ in the virtual corrections to $b \to s g$.
The $O_8$ contribution to the 
decay width $\G_8(b \to s f \overline{f})$ yields
\beq
\label{quarkdecay}
\G_8(b \to s f \overline{f}) = \frac{m_b^5 \, |G_F \, V_{ts}^*V_{tb} 
\, C_8^{0,\rm{eff}}|^2}{72\p^5} \, \a_s^2 \, \left[ \ln \frac{m_b}{2m_f} 
-\frac{2}{3} \right] \ .
\eeq 
%Comparing this result with $\G^{\rm{virt}}$ in eq. (\ref{gammavirt}),
%we see explicitly, that the mentioned logarithms indeed cancel.
  
\section{Gluon bremsstrahlung contributions}
\label{bremsmat}
\begin{figure}[t]
\begin{center}
\leavevmode
\includegraphics[width=\textwidth]{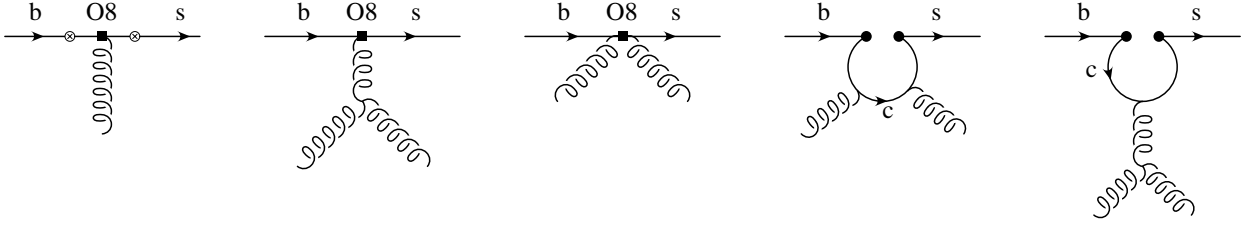}
\end{center}
\caption[f1]{Complete list of Feynman diagrams for $b \to s g g$, 
associated with the operators $O_1$, $O_2$ and $O_8$.}
\label{fig:3}
\end{figure}

In this section we discuss the gluon bremsstrahlung corrections 
to $b \to s g$, i.e. the process
$b \to s g g$. As in the case of the virtual corrections, 
we neglect contributions from the gluonic penguin operators 
as their Wilson coefficients are rather small. In this
approximation, the
matrix element $M^{\rm{brems}}(b \to s g g)$  
is of the form
\beq
\label{mbrems}
M^{\rm{brems}} = \frac{4G_F i}{\sqrt{2}} \, V_{ts}^* V_{tb} \, \left[
C_1^0 \, M_1^{\rm{brems}} +  C_2^0 \,M_2^{\rm{brems}} +
C_8^{0,\rm{eff}} \, M_8^{\rm{brems}} \right] \ ,
\eeq
where the three terms on the r.h.s. correspond to the contributions
of the operators $O_1$,  $O_2$ and $O_8$, respectively.
The corresponding Feynman diagrams are shown in fig. \ref{fig:3}. 
We note that in eq. (\ref{mbrems}) only the leading order pieces of the
Wilson coefficients are needed. 

The decay width $\G^{\rm{brems}}(b \to s g g)$ is then obtained by squaring 
$M^{\rm{brems}}$, followed by phase space integrations.  
These integrals are plagued with infrared and collinear 
singularities. Configurations with one gluon flying collinear to the
$s$-quark are regulated by a small strange quark mass $m_s$,
while 
configurations with two collinear gluons, or one soft gluon
are dimensionally regularized. 
As in the calculations of the virtual corrections, we
write the dimension as $d=4-2\e$. (Note that $\e$ has to be negative
in order to regulate the phase space integrals). 

When squaring $M^{\rm{brems}}$ in eq. (\ref{mbrems}), nine terms
are generated, which we denote for obvious reasons 
by $(O_1,O_1^*)$, $(O_1,O_2^*)$,
$(O_1,O_8^*)$, $(O_2,O_1^*)$, ...., $(O_8,O_8^*)$. 
We find that all these quantities 
are free of infrared and collinear singularities, except $(O_8,O_8^*)$.
Hence, one
can put $m_s=0$ in the finite terms and evaluate
the phase space integrals in $d=4$ dimensions. In the following,
we denote this finite contribution to the decay width by 
$\G^{\rm{brems}}_{\rm{fin}}$.
It turns out that only $\sim 5\%$ of the total NLL correction 
are due to $\G^{\rm{brems}}_{\rm{fin}}$.
As the analytical results for this finite piece,  written in terms
of  two-dimensional integrals, are rather lengthy, we skip the explicit 
expressions
in this letter; we stress, however, that 
$\G^{\rm{brems}}_{\rm{fin}}$, although small, will be retained in the
numerical evaluations.   

We now turn to the $(O_8,O_8^*)$ contribution, denoted 
by $\G^{\rm{brems}}_{88}$. 
After a lengthy, but straightforward calculation, 
we obtain ($\rho=(m_s/m_b)^2$;
$V=\a_s \, m_b^5 \, |G_F V_{ts}^*V_{tb}|^2/(24\p^4)$)
\beq
\label{brems88p}
\G^{\rm{brems}}_{\rm{88}} = 
\frac{  \a_s \left( C_8^{0,\rm{eff}} \right)^2 \, V \,
 \left( \frac{m_b}{\m} \right)^{-4\e}}{12 \, \p} \,
\, \left[ \frac{18}{\e^2_{\rm{IR}}} + \frac{67 + 8 \ln \rho}{\e_{\rm IR}} - 
4 \ln^2\rho+12 \ln\rho+ 240 - \frac{62\p^2}{3} 
\right] \ .
\eeq
The total decay with for $b \to s g g$ is then 
given by
\beq
\label{gammabremstot}
\G^{\rm{brems}}(b \to s g g ) =
\G^{\rm{brems}}_{\rm{fin}}+
\G^{\rm{brems}}_{88} \ .
\eeq
%
%%%%%%%%%%%%%%%%%%%%%%%%%%%%%%%%%%%%%%%%%%%%%%%%%%%%%
\section{Combined NLL branching ratio for $\boldsymbol{\symbol{98} \to
    \symbol{115}\symbol{103}}$ and $\boldsymbol{\symbol{98} \to \symbol{115}\symbol{103}\symbol{103}}$}
\label{brnll}
In this section we combine the decay widths for the virtually
corrected process
$b \to s g$ and the bremsstrahlung process $b \to s g g$ to the total
NLL decay width decay $\G^{\rm{NLL}}(b \to s g)$. 
We also absorbe in this quantity the $O_8$ induced contribution to the
processes $b \to s f \overline{f}$ ($f=u,d,s$), as discussed 
in section \ref{Quarkstrahlung}. 
%The expression
%for $\Gamma^{\rm{virt}}$, which contains the lowest order contribution
%to the decay width for $b \to s g$, together with its virtual corrections,
%may be found in eq. (\ref{gammavirt}). The result for the bremsstrahlung
%process, $\Gamma^{\rm{brems}}$ is given in eq. (\ref{gammabremstot}).
From the explicit formulas for $\Gamma^{\rm{virt}}$ and $\Gamma^{\rm{brems}}$
one can see that the infrared and collinear singularities 
cancel in the sum. 
The terms containing logarithms of the light quark masses $m_f$,
present in the result for $\Gamma^{\rm{virt}}$, are cancelled when combined
with  $\G_8(b \to s f \overline{f})$.
Putting together the individual pieces, the final result
for $\G^{\rm{NLL}}(b \to s g)$ can be written as
\beq
\label{rewrite}
\G^{\rm{NLL}}(b \to s g) =  
\frac{\a_s(m_b) \, m_b^5 }{24\p^4} |G_F V_{ts}^* V_{tb}|^2 \,
|\overline{D}|^2 + \G_{\rm{fin}}^{\rm{brems}} \ ,
\eeq
with
\bea
\label{dbar}
\overline{D}  &=& C_8^{0,\rm{eff}} + 
\frac{\a_s}{4\p} \, \left[
C_8^{1,\rm{eff}} 
- \frac{16}{3} C_8^{0,\rm{eff}}
+ C_1^0 [\ell_1 \, \log \frac{m_b}{\m} + r_1]  \right. \nn \\
& & \left. \quad \quad \quad
+ C_2^0 [\ell_2 \, \log \frac{m_b}{\m} +r_2]
+ C_8^{0,\rm{eff}} [(\ell_8+8+\b_0) \, \log \frac{m_b}{\m} + r_8] \right]  \ .
\eea
%In eq. (\ref{rewrite}), the quantity $\G_{\rm{fin}}^{\rm{brems}}$
%contains all the bremsstrahlung corrections except those originating
%from the $(O_8,O_8^*)$ interference, as dicussed earlier. 
A remark concerning the modulus square of the function $\overline{D}$
is in order: By construction, 
this square is understood to be taken in the same
way as the in the virtual contributions, i.e. by systematically
discarding the $O(\a_s^2)$ term. In this sense, the quantity
$\overline{D}$ can be viewed as an effective 
matrix element. We stress however
that, besides the virtual corrections, also 
the informations of 
$\G_{88}^{\rm{brems}}$ and 
$\G_8(b \to f \overline{f})$ are contained in the function $\overline{D}$. 

The quantities $r_1$ and $r_2$ appearing in eq. (\ref{dbar})
are given in eqs. (\ref{rer1ndr}) and (\ref{rer2ndr}), respectively.  
The explicit expressions for $\ell_1$, $\ell_2$, $\ell_8$ and $r_8$, 
read
\beq
\label{l8r8}
\ell_1=\frac{173}{162} \ ; \quad
\ell_2=\frac{70}{27} \ ; \quad
\ell_8=-\frac{19}{3} \ ; \quad
r_8=\frac{1}{18} \, \left[ 351 -19 \p^2 -36 \ln2 + 
6 \ln \frac{m_c^2}{m_b^2} \right] \ .
\eeq   
Note that all scale dependent quantities in eqs. 
(\ref{rewrite}) and (\ref{dbar}) 
are understood to be evaluated at the scale $\mu$, unless
indicated explicitly in the notation.

We would like to point out that $\ell_1$, $\ell_2$ and $(\ell_8+8+\b_0)$
are identical to the anomalous dimension matrix elements
$\g^{0,\rm{eff}}_{18}$,  
$\g^{0,\rm{eff}}_{28}$, and  $\g^{0,\rm{eff}}_{88}$, respectively,
which are given in ref. \cite{Misiak97}.
This is  what has to happen: Only in this case the
leading scale dependence of $C_8^{0,\rm{eff}}(\mu)$ gets compensated
by the second term in eq. (\ref{dbar}).

The NNL branching ratio ${\cal B}^{\rm{NLL}}(b \to s g)$ 
is then obtained as
\beq
\label{BRdef}
{\cal B}^{\rm{NLL}}(b \to s g) = 
\frac{\G^{\rm{NLL}}(b \to s g)}{\G_{\rm{sl}}} \,
{\cal B}_{\rm{sl}}^{\rm{exp}} \ ,
\eeq
where ${\cal B}_{\rm{sl}}^{\rm{exp}}$ denotes the experimental
semileptonic branching ratio of the $B$-meson. $\G_{\rm{sl}}$
stands for the theoretical expression of the semileptonic decay width of the
$B$-meson. Neglecting non-perturbative corrections of the order 
$(\L_{\rm{QCD}}/m_b)^2$, $\G_{\rm{sl}}$ reads (with $x_c=(m_c/m_b)$)
\beq
\label{gammasl}
\Gamma_{\rm{sl}} \approx \G(b \to c e \overline{\n}_e) =
\frac{G_F^2 \, m_b^5}{192\p^3} \, |V_{cb}|^2 \, g(x_c) \,
\left[1 + \frac{\a_s(\mu)}{2\p} \, h_{\rm{sl}}(x_c) + O(\a_s^2) \right] \ ,
\eeq
with the phase space function 
$g(x_c) = 1 - 8 \, x_c^2 - 24 \, x_c^4 \, \ln x_c + 8 \, x_c^6 - x_c^8 $.
The analytic expression for $h_{\rm{sl}}(x_c)$ can be found in ref.
\cite{Nir}. The approximation (taken from ref. \cite{Lenz1})
\beq
h_{\rm{sl}}(x_c)=-3.341 + 4.05 \, (x_c-0.3) - 4.3 \, (x_c-0.3)^2 \ , 
\eeq
which we use in the following, 
holds to an accuracy of one permille for
$0.2 \le x_c \le 0.4$.
%We note that in the numerical analysis we systematically expand 
%the expression for the branching ratio (\ref{BRdef}) in $\a_s$, dropping
%terms of $O(\a_s^2)$. 
%Also a short remark 
%concering the LL branching ratio is in order: For the
%decay width $\G^{\rm{LL}}(b \to s g)$, we use the expression
%\beq
%\G^{\rm{LL}}(b \to s g) =  
%\frac{\a_s(\mu) \, m_b^5 }{24\p^4} |G_F V_{ts}^* V_{tb}|^2 \,
%\left( C_8^{LL,\rm{eff}}(\mu)\right)^2 \ .
%\eeq
%The LL branching ratio for $b \to s g$ is then obtained as in eq. 
%(\ref{BRdef}), but by discarding the radiative corrections in 
%$\G_{\rm{sl}}$.  

\section{Numerical results for the combined branching ratio}
\label{numres}
\begin{figure}[t]
\begin{center}
\leavevmode
\includegraphics[width=8.0cm]{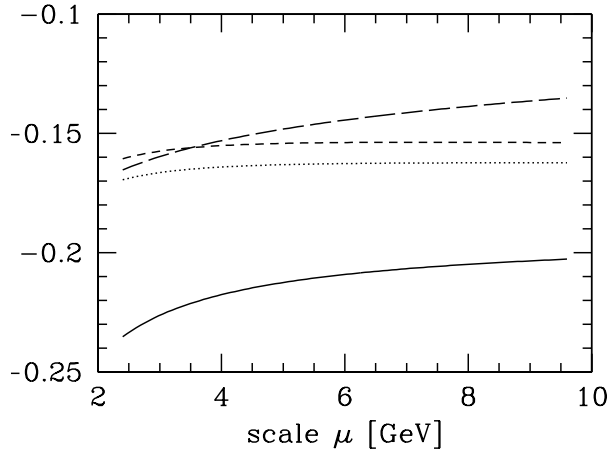}
\vspace{3ex}
\caption[f1]{Scale ($\mu$) dependence of the function $\overline{D}$
(see eq. (\ref{dbar})) in various approximations: The long-dashed line shows
$C_8^{0,\rm{eff}}$; the short-dashed line corresponds to putting
$r_1=r_2=r_8=0$; the dotted line is obtained by only putting $r_2=0$;
the solid line shows the full function $\overline{D}$. See text.}
\label{fig:dfun}
\end{center}
\end{figure}
We first discuss the sizes of the various
NLL corrections at the level of the function $\overline{D}$, defined in
eq. (\ref{dbar}).
As  already stated, the terms containing explicit
logarithms of the ratio $(m_b/\m)$ get compensated by the scale dependence
of the first term on the r.h.s. of eq. (\ref{dbar}). This feature
can be observed in fig. \ref{fig:dfun}, when comparing the two dashed
lines. The long-dashed line represents only the first term $C_8^{0}$
of the function $\overline{D}$, while the short-dashed
line shows $\overline{D}$, in which  $r_1$, $r_2$ and $r_8$
are put to zero. As expected, the short-dashed line has a milder 
$\mu$-dependence. When switching on also $r_1$ and $r_8$ 
(but keeping $r_2=0$), the resulting curve, shown by the dotted line,
stays close to the short-dashed curve and the scale dependence remains
very mild. However, when switching on also $r_2$, the situation changes
drastically. The resulting solid line, which  represents the full NLL 
$\overline{D}$ function, implies that the term containing
the two-loop quantity $r_2$, induces a large NLL correction.
As this large correction term contains a factor $\a_s(\mu)\, C_2(\mu)$,  
it is of no surprise, that the NLL prediction for the function 
$\overline{D}$ suffers from a relatively large scale dependence, as illustrated
by the solid line.

\begin{figure}[t]
\begin{center}
\leavevmode
\includegraphics[width=8.0cm]{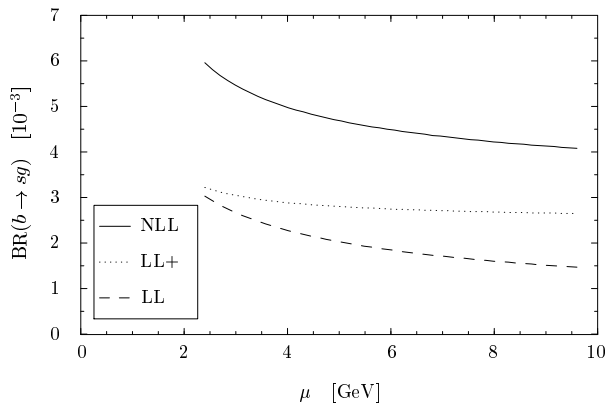}
\vspace{3ex}
\caption[f1]{Branching ratio ${\cal B}(b \to s g)$ as a function
of the scale $\mu$ in various approximations: The dashed and the solid
lines show the LL and the NLL predictions, respectively; the dotted
line is obtained by putting $r_1=r_2=r_8=\G_{\rm{brems}}^{\rm{fin}}=0$
in the NLL expression for $\G^{\rm{NLL}}(b \to s g)$ in 
eq. (\ref{rewrite}). See text.}
\label{fig:br}
\end{center}
\end{figure}

The NLL branching ratio ${\cal B}^{\rm{NLL}}(b \to s g)$ is then obtained
as described in section \ref{brnll}. The result is shown by the solid
line in fig. \ref{fig:br}. For the  input values, we choose:
$m_b=(4.8 \pm 0.2)$ GeV, $(m_c/m_b)=(0.29 \pm 0.02)$, 
$\a_s(m_Z) = 0.119 \pm 0.003$, $|V_{ts}^* V_{tb}/V_{cb}|^2=0.95 \pm 0.03$, 
${\cal B}_{\rm{sl}}^{\rm{exp}}=(10.49 \pm 0.0046)\%$, and $m_t^{\rm{pole}}=
(175 \pm 5)$ GeV.  
As the scale dependence is rather large, we did not take
into account the error due to the uncertainties in the input parameters. 
Based on fig. \ref{fig:br}, we obtain the NLL branching ratio
\beq
\label{brvalue_nll}
{\cal B}^{\rm{NLL}}(b \to s g) = (5.0 \pm 1.0) \times 10^{-3} ,
\eeq
which is more than a factor two larger than
the LL value
\beq
\label{brvalue_ll}
{\cal B}^{\rm{LL}}(b \to s g) = (2.2 \pm 0.8) \times 10^{-3} \ ,
\eeq
extracted from the dashed line in fig. \ref{fig:br}.
As stressed in the discussion of the function $\overline{D}$,
the main enhancement is due to the virtual- and bremsstrahlung
corrections to $b \to s g$, calculated in this paper.
At the level of the branching ratio, this fact is illustrated by 
the dotted line in fig. \ref{fig:br}, 
which is obtained by discarding $\G_{\rm{fin}}^{\rm{brems}}$
and by switching off $r_1$, $r_2$ and $r_8$ in  the expression
for $\G^{\rm{NLL}}(b \to s g)$ (see eq. (\ref{rewrite})).

The largest uncertainty due to the physical input parameters
(other than $\mu$) 
on ${\cal B}(b \to s g)$ results from the charm quark mass.
Varying $x_c=m_c/m_b$ between 0.27 and
0.31 and choosing $\mu=m_b$, the resulting uncertainty amounts to
$\sim \pm 6 \%$.

\vspace{0.5cm}
\noindent

{\bf Acknowledgments}:
We would like to thank A. Ali, A. Kagan, A. Lenz, P. Minkowski, M. Neubert,
and U. Nierste for helpful discussions.

\end{document}